\documentclass[numberedappendix]{emulateapj}
\usepackage{amssymb,amsmath}
\usepackage{verbatim}
\usepackage{color,hyperref}
\usepackage{needspace}
\usepackage{enumitem}
\definecolor{linkcolor}{rgb}{0,0,0.25}
\hypersetup{
  colorlinks=true,        
  linkcolor=linkcolor,    
  citecolor=linkcolor,    
  filecolor=linkcolor,    
  urlcolor=linkcolor      
}
\newcounter{address}
\newcommand{\etal}{et al.}

\newcommand{\eg}{e.g.}
\newcommand{\eqnname}{equation}

\newcommand{\equationname}{\eqnname}

\renewcommand{\figurename}{Figure}

\newcommand{\sectionname}{$\mathsection$}

\newcommand{\degp}{\(\stackrel{\circ}{\textstyle.\rule{0pt}{0.65ex}}\)}

\newcommand{\yr}{\ensuremath{\,\mathrm{yr}}}
\newcommand{\Myr}{\ensuremath{\,\mathrm{Myr}}}
\newcommand{\Gyr}{\ensuremath{\,\mathrm{Gyr}}}
\newcommand{\kpc}{\ensuremath{\,\mathrm{kpc}}}
\newcommand{\pc}{\ensuremath{\,\mathrm{pc}}}
\newcommand{\kms}{\ensuremath{\,\mathrm{km\ s}^{-1}}}
\newcommand{\msun}{\ensuremath{\,\mathrm{M}_{\odot}}}
\newcommand{\mas}{\ensuremath{\,\mathrm{mas}}}

\newcommand{\inv}{\ensuremath{^{-1}}}

\submitted{}

\begin{document}

\title{The shape of the inner Milky Way halo from observations of the Pal 5 and GD--1 stellar streams}

\author{Jo~Bovy\altaffilmark{1,2}, Anita Bahmanyar\altaffilmark{1}, Tobias~K.~Fritz\altaffilmark{3}, and Nitya~Kallivayalil\altaffilmark{3}}
\altaffiltext{\theaddress}{\label{1}\stepcounter{address}
  Department of Astronomy and Astrophysics, University of Toronto, 50
  St.  George Street, Toronto, ON, M5S 3H4, Canada;
  bovy@astro.utoronto.ca~}
\altaffiltext{\theaddress}{\label{2}\stepcounter{address}
  Alfred~P.~Sloan~Fellow}
\altaffiltext{\theaddress}{\label{3}\stepcounter{address}
  Department of Astronomy, University of Virginia, Charlottesville,
3530 McCormick Road, VA 22904-4325, USA}

\begin{abstract}
  We constrain the shape of the Milky Way's halo by dynamical modeling
  of the observed phase--space tracks of the Pal 5 and GD--1 tidal
  streams. We find that the only information about the potential
  gleaned from the tracks of these streams are precise measurements of
  the shape of the gravitational potential---the ratio of vertical to
  radial acceleration---at the location of the streams, with weaker
  constraints on the radial and vertical accelerations separately. The
  latter will improve significantly with precise proper-motion
  measurements from \emph{Gaia}. We measure that the overall potential
  flattening is $0.95\pm0.04$ at the location of GD--1 ($[R,Z] \approx
  [12.5,6.7]\kpc$) and $0.94\pm0.05$ at the position of Pal 5 ($[R,Z]
  \approx [8.4,16.8]\kpc$). Combined with constraints on the force
  field near the Galactic disk, we determine that the axis ratio of
  the dark--matter halo's density distribution is $1.05\pm0.14$ within
  the inner 20 kpc, with a hint that the halo becomes more flattened
  near the edge of this volume. The halo mass within $20\kpc$ is
  $1.1\pm0.1\times10^{11}\msun$. A dark--matter halo this close to
  spherical is in tension with the predictions from numerical
  simulations of the formation of dark--matter halos.
\end{abstract}


\section{Introduction}

The Milky Way's gravitational potential within the inner tens of kpc
is key to various important problems in galactic astrophysics and
near-field cosmology. Containing the vast majority of all of its
stars, hosting almost all of its building blocks, and being amenable
to detailed observational characterization, this volume contains a
wealth of information on the formation of the Galactic bulge and disk
deep in the Milky Way's potential well, the formation channels of the
stellar halo, and the detailed large- and small-scale structure of the
dark matter halo and its response to the growth of the baryonic
components. It is in this volume---soon to be mapped in detail by the
\emph{Gaia} satellite \citep{deBruijne12a}---that we can determine the
Milky Way's mass and light budget
\citep[\eg,][]{Flynn06a,Bovy13a,McKee15a}, the dark--matter halo's
shape and radial profile near and within its scale radius
\citep[\eg,][]{Olling00a,Binney15a}, and the abundance of low-mass
substructure \citep[\eg,][]{Johnston02a,Ibata02a}. Determining the
three-dimensional force field is a prerequisite to any study of the
orbital properties of stars and star clusters in the disk and halo
\citep{Binney13a,Rix13a}.

The shape of the dark--matter halo is of particular interest and
poorly constrained by current data. Because the primary constraint on
the gravitational potential in the Milky Way and external galaxies
alike has been the rotation curve, which is hardly sensitive to the
detailed three-dimensional shape of a mass distribution, precise
measurements of the shapes of galactic dark--matter halos are close to
non-existent. This is an unfortunate situation, because numerical
simulations of the formation of dark--matter halos make strong
predictions for their three-dimensional shapes. In the absence of
baryons, dark--matter halos are generally found to be strongly
triaxial
\citep[\eg,][]{Frenk88a,Dubinski91a,Warren92a,Cole96a,Jing02a,VeraCiro11a}. The
growth of a galactic baryonic disk causes the halos to become more
axisymmetric and aligned with the disk, but the minor-to-major axis
ratio $c/a$ changes only by a few tenths to $c/a \approx 0.7$ to $0.8$
\citep{Dubinski94a,Gustafsson06a,Debattista08a,Abadi10a,Kazantzidis10a}. The
amount of sphericalization depends on the mass of the baryonic
component \citep{Kazantzidis10a}. Because the Milky Way's disk is
close to maximal \citep{Bovy13a,Wegg16a}, the inner dark--matter halo
is expected to have $c/a \approx0.8$. This is an especially
interesting prediction, as, \eg, non-canonical dark--matter models
such as self-interacting dark matter predict that halos should be
spherical in their high-density inner regions
\citep{Spergel00a,Yoshida00a,Dave01a,Miralda02a,Peter13a}.

In the Milky Way we can measure the three-dimensional force (or,
equivalently, acceleration) field and thus the halo's shape in
different ways. Assuming that the inner stellar halo is in a
quasi-stationary state, equilibrium modeling using the Jeans equations
\citep[\eg,][]{Loebman14a,Bowden16a} or using the Jeans theorem
\citep[\eg,][]{Binney15a} can be used to determine the force field
from the observed positions and velocities of halo stars. So far, this
approach has had limited success due to the difficulty of observing
the stellar halo's kinematics over a large enough volume and at high
enough precision. Measurements of the shape of the inner halo ($r
\lesssim 20\kpc$) from these measurements range from strongly oblate
($c/a \approx 0.4$; \citealt{Loebman14a}) to prolate
\citep{Bowden16a}. The trajectories of hypervelocity stars are also
sensitive to the shape of the force field and can in the future be
used to constrain the dark--matter halo's shape \citep{Gnedin05a}.

The current best measurements of the shape of the Milky Way's
gravitational potential in the halo region come from observations of
stellar streams. Streams form when a satellite galaxy or a globular
cluster gets tidally destroyed in the Milky Way's gravitational
potential \citep{Johnston98a,Helmi99a,Tremaine99a}: mass loss due to
tidal stripping leads to leading and trailing arms at slightly
different orbital energies than the progenitor cluster. The track
traced by the stellar stream is close to an orbit
\citep{Eyre11a,Sanders13a} and thus provides a rather direct
measurement of the local acceleration near the stream. Dynamical
modeling of tidal streams has made much progress in the last few
years, with multiple distinct methods proposed for realistically
approximating the formation and evolution of streams in general
gravitational potentials
\citep[\eg,][]{Eyre11a,Sanders13b,Varghese11a,Bovy14a,PriceWhelan14a,Sanders14a,Amorisco15a,Fardal15a,Kuepper15a}. However,
this progress in modeling has led to only a modest improvement in the
observational modeling of tidal streams and determining the shape of
the halo. Variants of particle-spray-type modeling, where the
formation of the stream is simply modeled by orbit integration of a
finite number of ejected stars, have been used to constrain the
potential shape using the GD--1 \citep{Grillmair06a} and Pal 5 streams
\citep{Odenkirchen01a}. \citet{Bowden15a} used the data from
\citet{Koposov10a} to determine the overall potential flattening
$q_\Phi$ to be $q_\Phi = 0.90^{+0.05}_{-0.10}$ in a simple logarithmic
potential model. \citet{Kuepper15a} employed a three-component
disk--bulge--halo model to determine the shape of the halo's
potential, finding $q^h_\Phi \approx 0.95^{+0.15}_{-0.12}$, but fixing
the parameters describing the disk and bulge. The uncertainties in
these measurements do not allow a stringent test with respect to the
predictions from numerical simulations discussed above.

A second class of approaches uses action--angle coordinates to build
simple models of tidal streams
\citep{Helmi99a,Tremaine99a,Eyre11a,Bovy14a,Sanders14a}. In
particular, the simple modeling approach proposed in \citet{Bovy14a}
(hereafter B14) and \citet{Sanders14a} allows one to make smooth,
continuous predictions for the present-day structure of an observed
stream. In this paper we present the first application of this
technique to observational data, by modeling the Pal 5 and GD--1
streams. We use general models for the Milky Way's gravitational
potential with a large amount of freedom to investigate the exact
nature of the constraints on the potential provided by these
streams. We demonstrate that even with the current data, these streams
are exquisitely sensitive to the shape of the dark--matter halo and
the overall shape of the potential.

The structure of this paper is as follows. We discuss our Milky-Way
model and prior constraints on the shape of the dark--matter halo from
existing data in \sectionname~\ref{sec:prior}. In
\sectionname~\ref{sec:method} we present the details of our
stream-fitting methodology. We then apply this methodology to
observational data for Pal 5 in \sectionname~\ref{sec:pal5} and for
GD--1 in \sectionname~\ref{sec:gd1}, including comparisons to previous
fits in the literature. We combine our new measurements of the
force-field near Pal 5 and GD--1 with constraints coming primarily
from the rotation curve and vertical disk dynamics in
\sectionname~\ref{sec:combined} to make a determination of the axis
ratio $c/a$ of the dark--matter halo's density distribution and
discuss these results. We present our conclusions in
\sectionname~\ref{sec:conclusion}. 

\section{Milky Way model and prior constraints}\label{sec:prior}

\subsection{Milky Way model}\label{sec:mwmodel}

For most of the analyses in this paper we use a simple three-component
potential model for the Milky Way consisting of a bulge, disk, and
halo component. The bulge is in all cases modeled with a spherical
power-law density profile that is exponentially cut-off. The power-law
exponent is fixed to $-1.8$ and the exponential cut-off radius is set
to 1.9 kpc. In most cases, we model the disk component as a
Miyamoto-Nagai potential with three free parameters: the scale length
$h_R$, the scale height $h_z$, and an amplitude parameter. We will
also explore modeling the disk as a double exponential density profile
in the radial and vertical directions, which has the same three free
parameters.

\begin{figure*}
  \includegraphics[width=\textwidth,clip=]{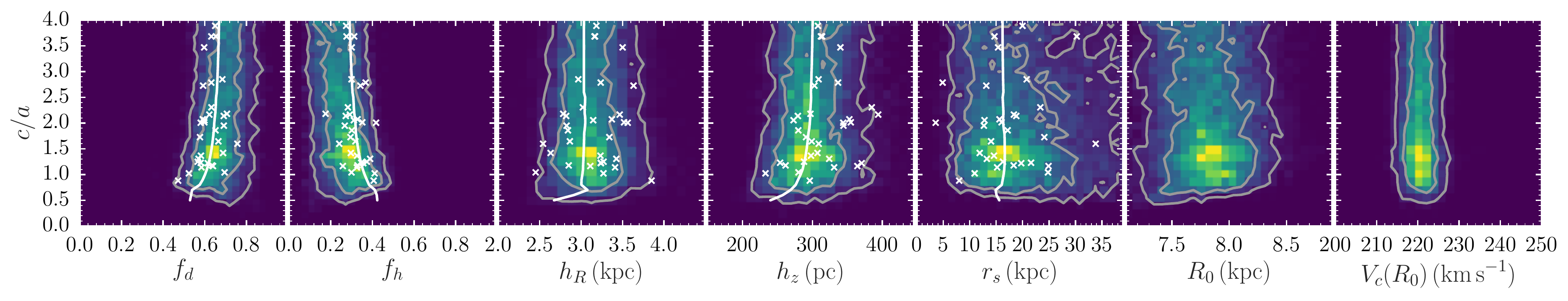}
  \caption{Constraints on the halo-axis ratio $c/a$ and the disk,
    bulge, and halo components of the Milky Way from dynamical data
    prior to the stream measurements from this paper. This figure
    displays the PDF of the eight parameters describing the Milky Way
    potential, focusing on the correlation between $c/a$ and the other
    parameters. The contours contain $68\,\%$ and $95\,\%$ of the
    distribution. The white line is the locus of the best-fit values
    of the parameters shown horizontally as a function of $c/a$ for
    $R_0 = 8\kpc$ and $V_c(R_0) = 220\kms$. The white crosses display
    the values of the fixed parameters for the 32 potential families
    used in the analysis of Pal 5 and GD--1; only the horizontal
    parameters are fixed in these families. The dynamical data prior
    to the Pal 5 and GD--1 stream data do not strongly constrain the
    halo axis ratio, except to rule out strongly flattened halo
    densities with $c/a \lesssim0.7$ (2.5\,\%
    confidence).}\label{fig:mwpot14_prior}
\end{figure*}

The halo component is represented by a triaxial Navarro-Frenk-White
(NFW) profile \citep{Navarro97a} with constant density along similar,
co-axial ellipsoids. In the frame $(x',y',z')$ aligned with the three
symmetry axes we have that
\begin{equation}\label{eq:nfw}
\rho^h(x',y',z') = \frac{M}{4\,\pi\,r_s^3}\,\frac{1}{(m/r_s)\,(1+m/r_s)^{2}}\,,
\end{equation}
where
\begin{equation}
  m^2 = x'^2 + \frac{y'^2}{(b/a)^2}+\frac{z'^2}{(c/a)^2}\,.
\end{equation}
We also allow the orientation of the $(x',y')$ frame to be specified
with respect to the Sun--Galactic-center line ($Y=0$ in the
Galactocentric frame $(X,Y,Z)$), but keep $z'\equiv Z$. The axis ratio
$c/a$ then describes the flattening of the halo perpendicular to the
Galactic disk and $b/a$ deviations from axisymmetry in the disk
plane. We compute the gravitational potential and forces corresponding
to this density using the expressions in \citet{Merritt96a},
performing the necessary integrals with 50-th order Gauss-Legendre
quadrature. This potential--density pair as well as the generalization
to arbitrary orientations between $(X,Y,Z)$ and $(x',y',z')$ has been
implemented into \texttt{galpy} \citep{Bovy15a} for the purpose of
this work.

There are few robust constraints on the axis ratio $b/a$ of the
dark--matter halo in the Milky Way. If $b/a \neq 1$ in the inner Milky
Way, the closed orbits in the disk around which stellar orbits librate
would not be circular. This non-circularity can be detected using the
chemical and kinematic properties of disk stars. The lack of azimuthal
abundance differences in the few kpc near the Sun constrains $b/a$ to
be near one or $y$ and $y'$ to be close to aligned
\citep{Bovy14b}. The kinematics of disk stars within a few kpc from
the Sun strongly constraints $b/a$ to be close to one for any angle
between $y$ and $y'$ \citep{Bovy15b}.

Other constraints on $b/a$ come from tidal streams. A successful model
of the tidal tails of the Sgr dwarf galaxy has been found with $b/a
\approx 0.7$ \citep{Law10a}, but if this axis ratio is assumed in the
inner $\approx 20\kpc$, no model compatible with the observed Pal 5
tidal stream can be found \citep{Pearson15a}.

We can use the sensitivity to $b/a$ and $c/a$ of orbit segments
compatible with the Pal 5 and GD--1 data described below as a proxy
for how sensitive these data on the average location of the track are
to these parameters. For both streams we find that such orbit segments
are hardly sensitive to even large variations in $b/a$ ($b/a$ going
from 1/2 to 2), whether or not $(X,Y)$ and $(x',y')$ are aligned or
not. The sensitivity to the maximally misaligned setup is for both
streams larger than that of the aligned setup, but all changes in
$b/a$ lead to orbit segments that differ by an order of magnitude less
than orbits for different $c/a$. This is the case because the streams
are located near a Galactocentric azimuth of zero (GD--1) and $\pi$
(Pal 5), that is, both are approximately along the line connecting the
Sun to the Galactic center.

The lack of sensitivity to changes in $b/a$ only apply to the stream
\emph{track}. Like \citet{Pearson15a}, we find that if the halo is
triaxial with $b/a \neq 1$ a narrow stream does not form at the
phase--space location of Pal 5 because the orbital history of the
stream is chaotic. Even though the orbit segments near the stream do
not depend strongly on $b/a$, orbits compatible with the stream
integrated for multiple Gyr are chaotic for $b/a \neq 1$ and the
stream would therefore not be narrow. This fact could be used to set a
limit on $|b/a-1|$ from the observed width of Pal 5, but we do not
pursue this in this paper.

Because observational data indicate that $b/a \approx 1$ and because
both the Pal 5 and GD--1 stream tracks are largely insensitive to
changes in $b/a$, we fix $b/a$ to one in what follows.

\subsection{Prior constraints on the halo axis ratio}\label{sec:priorfit}

To determine what, if any, constraints on the halo axis ratio $c/a$ we
can set based on existing dynamical data and to find a realistic set
of potentials to use in the analysis of the Pal 5 and GD--1 streams
below, we perform fits of the above potential model to a variety of
dynamical data on the bulge, disk, and halo. This fit is similar to
that used to arrive at the fiducial \texttt{MWPotential2014} model of
\citet{Bovy15a} and we summarize the data, fit procedure, and
differences with \texttt{MWPotential2014} here.

The data are similar to those used by \citet{Bovy13a} and
\citet{Bovy15a}:
\begin{enumerate}[label=\emph{\alph*)}]
\item The velocity dispersion $\sigma_b = 117\pm15\kms$ measured in
  Baade's window \citep{Dehnen98a,binneytremaine};
\item The vertical force $|F_Z| = 67\pm6\,(2\pi G\,M_\odot\pc^{-2})$
  at the solar circle at $|z| = 1.1\kpc$ and the local visible surface
  density $\Sigma = 55\pm5\,M_\odot\pc^{-2}$ from \citet{Zhang13a};
\item The vertical force measurements at $|z| = 1.1\kpc$ of
  \citet{Bovy13a} (their Table 3);
\item The terminal-velocity measurements of \citet{Clemens85a} and
  \citet{McClure07a}, modeled in the same way as in
  \citet{Bovy13a}. In \sectionname~\ref{sec:combined} we replace the
  older \citet{Clemens85a} by the newer \citet{McClure16a} data;
\item The mid-plane ($Z=0$) density at the solar circle of
  \citet{Holmberg00a}: $\rho(R_0,Z=0) = 0.10\pm0.01\,M_\odot\pc^{-3}$;
\item The measurements of the logarithmic slope of the rotation curve
  from \citet{Bovy12a}, represented in the same way as in equation~(41)
  in \citet{Bovy13a};
\item The measurement of the total mass within $60\kpc$ from
  \citet{Xue08a}: $M(r<60\kpc) = 4.0\pm0.7\times10^{11}\,M_\odot$.
\end{enumerate}
Unlike \citet{Bovy15a}, we allow the Sun's distance to the Galactic
center $R_0$ and the circular velocity $V_c(R_0)$ at $R_0$ to
vary. Based on recent measurements of $R_0$
\citep{Chatzopoulos15a,BlandHawthorn16a,Boehle16a}, we use the
constraint $R_0 = 8.1\pm0.1\kpc$. We also use the constraint $V_c(R_0)
= 218\pm10\kms$ \citep{Bovy12a}. We will see that all determinations
of $c/a$ in this paper are uncorrelated with $R_0$ and $V_c(R_0)$, so
these constraints are not crucial.

The free parameters of the potential model are the amplitude, scale
length $h_R$, and scale height $h_z$ of the disk component, the
amplitude, scale radius $r_s$, and axis ratio $c/a$ of the halo
component, $R_0$, and $V_c(R_0)$. The amplitudes of the disk, halo,
and bulge components are specified in terms of the fraction of the
radial force at $(R,Z) = (R_0,0)$ that they provide (with $f_d$ for
the disk fraction and $f_h$ for the halo fraction). The results from
fitting the data above are summarized in
\figurename~\ref{fig:mwpot14_prior}. This figure displays the
posterior probability distribution function (PDF) when modeling the
disk using a Miyamoto-Nagai density, focusing on the correlation
between $c/a$ and the other model parameters. The PDFs are similar
when using a double-exponential disk, except that the scale length of
the disk is $h_R \approx 2.4\kpc$ (and $h_R \approx 2.2\kpc$ when also
including a gas component as discussed below, in agreement with
\citealt{Bovy13a}).

It is clear from \figurename~\ref{fig:mwpot14_prior} that there is
only a weak constraint on $c/a$ from these data and that $c/a$ is
largely uncorrelated with all of the other model parameters. However,
the data disfavor strongly flattened halo densities, with only
$2.5\,\%$ probability that $c/a < 0.7$ in the case of a Miyamoto-Nagai
disk and that $c/a < 0.8$ for a double-exponential disk (both of these
$c/a$ limits are about 0.2 smaller for $0.5\,\%$ probability). Note
that these models do not include a more extended gas
component. Including such a component to the double-exponential disk
model with a scale length that is twice that of the stellar component,
a scale height of $150\pc$, and a local surface density of
$10\msun\pc^{-2}$, the $2.5\,\%$ lower limit is 0.65. The lower
cut-off in $c/a$ in all of these models happens because the $F_Z$ data
from \citet{Bovy13a} indicate a short scale length ($\approx2.5\kpc$)
for the total mass within $\approx1\kpc$ of the disk mid-plane. The
halo's effective scale length is $\gtrsim 4\kpc$, thus making it
difficult to satisfy the short mass scale length implied by the $F_Z$
measurements if too much of the halo's mass is concentrated near the
mid-plane due to the halo's flattening. Above this lower cut-off on
$c/a$, the PDF for $c/a$ is almost flat up to $c/a = 4$, which is the
highest $c/a$ that we consider.

Thus, the pre-stream dynamical data do not provide a strong constraint
on $c/a$ beyond that it cannot be much smaller than one. We re-do this
fit in \sectionname~\ref{sec:combined} adding in the new measurements
of the force field from the next sections.

\section{Stream-fitting methodology}\label{sec:method}

The data on the Pal 5 and GD--1 stellar streams that we consider in
the next two sections come in the form of measurements of the
phase--space location of the track of these streams. That is, we have
measurements of some combination of the sky position, distance from
the Sun, proper motion vector, and line-of-sight velocity measured as
a function of a coordinate going along the stream. For the Pal 5
stream we also have measurements of the six-dimensional position and
velocity of its progenitor, the Pal 5 globular cluster. We do not make
use of measurements of the density along the stream.

To fit these data we follow the action--angle modeling approach of B14
(see also \citealt{Sanders14a}). The approach of B14 consists of a
simple analytic model for the frequency distribution of tidal debris
at the time of stripping. In action--angle coordinates, this analytic
model can be straightforwardly manipulated using the simple linear
action--angle dynamics to derive the current frequency--angle
structure along the stream for any given model. In particular, one can
easily and analytically compute the track in frequency and angle space
as a function of the location along the stream. \citet{Bovy14a} also
introduced a novel, general method for computing actions, frequencies,
and angles in any static gravitational potential (including triaxial
potentials) that can be used to convert this track in frequency--angle
space to configuration space. The track can also be converted to
configuration space in a more direct manner using the Torus Mapper
code of \citet{Binney16a}, but we primarily use the method of B14,
because we were unable to get the Torus Mapper to return action--angle
coordinates that were as accurate as those obtained using the B14
method. We only use the Torus Mapper in the rare instances where the
B14 method fails to converge. In configuration space, the track can be
compared to observed data. The fundamental ingredients of the model
are (a) a prescription for the times at which stars are stripped from
the progenitor, (b) a model for the distribution of frequency (and
angle) offsets from the progenitor at the time of stripping, (c) the
phase--space location of the progenitor, and (d) the gravitational
potential of the host galaxy (the Milky Way in this case). In the case
of Pal 5, ingredient (c) is strongly constrained by the measurements
of the position and velocity of the Pal 5 cluster. For GD--1, for
which the progenitor is unknown, the phase--space location of the
progenitor needs to be fully constrained by the stream data.

\begin{figure*}
  \includegraphics[width=\textwidth,clip=]{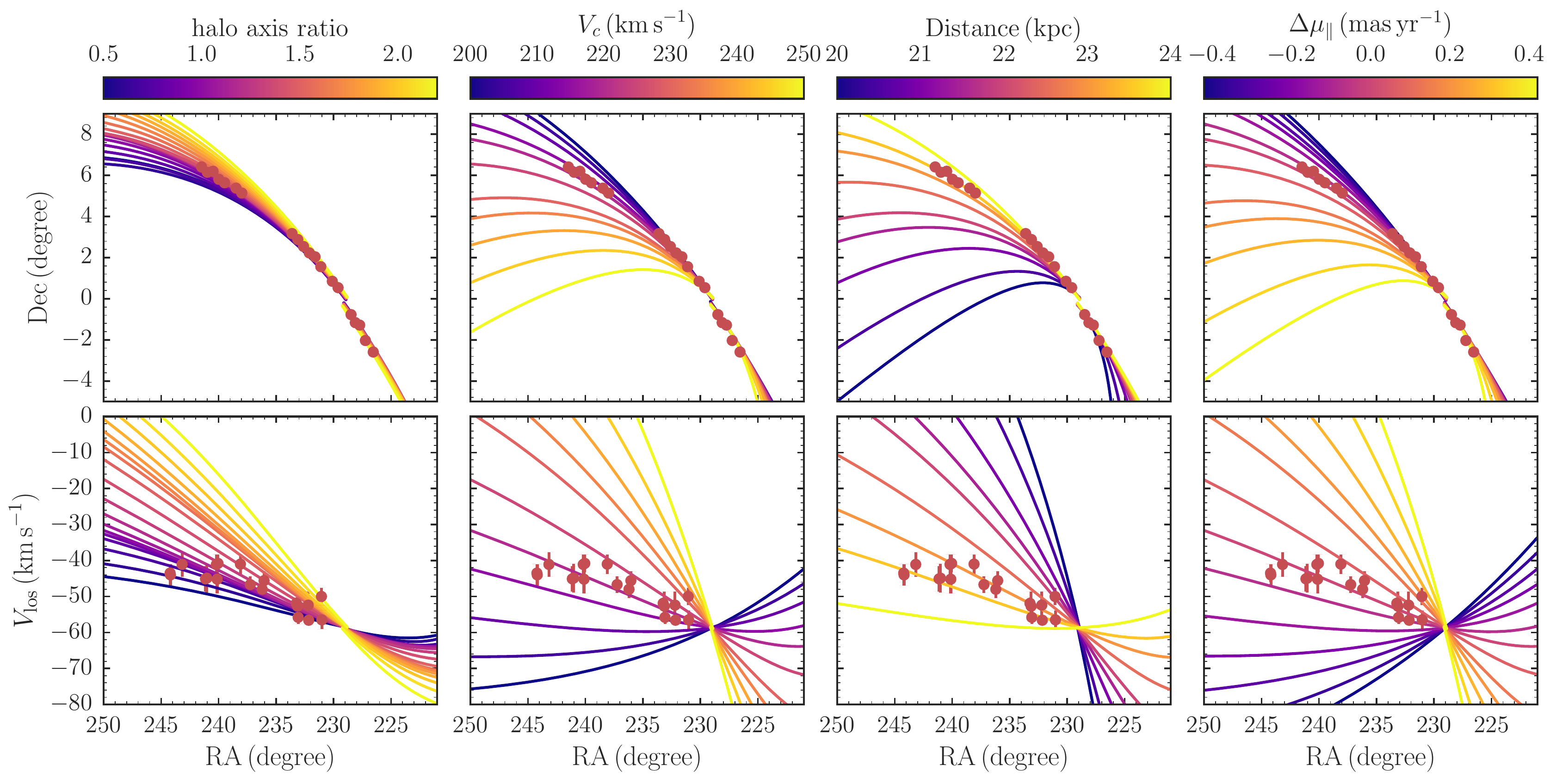}
  \caption{Sensitivity of the Pal 5 stream location to variations in
    the gravitational potential and in the position and proper motion
    of the Pal 5 progenitor cluster. The top panels display the
    location of the stream on the sky and the bottom panels give the
    line-of-sight velocity along the stream and include the data that
    we fit (see text). The leftmost column displays variations in the
    axis ratio $c/a$ of the halo, which changes the relative magnitude
    of the radial and vertical components of the force at Pal 5's
    location. The second column shows variations in the circular
    velocity $V_c$ at $R_0$, which changes the magnitude of the force,
    while keeping the ratio of the vertical-to-radial components the
    same. The third and fourth columns demonstrate the effect of
    changing the distance and the parallel component of the proper
    motion of the Pal 5 progenitor cluster. The overall normalization
    of the force (through $V_c$), the distance, and the proper motion
    all affect the location of the track in sky position and
    line-of-sight velocity in the opposite way (increasing Dec while
    decreasing the velocity at a given RA) and they are therefore to a
    large extent degenerate with each other. However, varying the axis
    ratio changes the track in both coordinates in the same direction
    and is therefore uncorrelated with the other parameters. The Pal 5
    stream is exquisitely sensitive to the halo's axis
    ratio.}\label{fig:pal5track}
\end{figure*}

B14 proposed a simple model for ingredients (a) and (b), in which the
stripping rate is constant up to a time $t_d$ in the past at which the
disruption started and in which the frequency offset distribution is
determined by a single velocity-dispersion parameter $\sigma_v$. B14
argued and demonstrated for a single $N$-body simulation of the
disruption of a GD--1-like cluster that the stripping rate can be
approximated as constant on timescales longer than the radial period;
on shorter timescales the stripping is dominated by pericentric
bursts. We have performed similar additional simulations in the
three-component Milky-Way-like potential \texttt{MWPotential2014} and
for clusters that are more concentrated than the loosely-bound cluster
used by B14 and find the same behavior in all cases. Furthermore, we
have computed the effect on the track of assuming a stripping rate
that decreases or increases in time. In all of these cases, the mean
frequency offset between the stream and the progenitor changes by less
than 5\,\% or less than 1/3 of the dispersion around the mean
frequency. Thus, the track is largely insensitive to changes in the
stripping rate and we assume that it is constant throughout the
analysis.

B14 further demonstrated that the mean frequency vector along a stream
is constant up to the edge of the stream, which can only be reached by
stars stripped at exceptionally high frequency differences (their
Figure 5). It is straightforward to compute the density along the
stream in the B14 model in a similar manner as the calculation of the
mean frequency along the stream and such calculations demonstrate that
the edge of the stream has much lower density than the part closer to
the progenitor. Thus, the segment of the stream that is high enough in
surface brightness to be detected observationally has an approximately
constant frequency vector. Therefore, we can model this part of the
stream as a single orbital torus. Note that this does \emph{not} mean
that we can model it as a single \emph{orbit}, because the direction
of the frequency offset between the progenitor and the stream is in
general not aligned with the direction of the frequency vector of an
orbit. Therefore, the path along the torus along which the stream lies
is not aligned with an orbit on the torus. This is especially the case
in the three-component Milky Way models and the non-spherical halos
that we consider here. We do not know the extent of either the Pal 5
or GD--1 stellar streams because they are limited by the edge of the
surveys in which they have been detected or by the Galactic
plane. Therefore, constraining the time $t_d$ at which disruption
started is difficult with the present data. We therefore fix $t_d$ to
$10\Gyr$ in both cases, making the stream old enough that for any
potential that we consider a long stream forms. If better measurements
of the width and length of streams were available, these could be used
as additional constraints, because in very non-spherical halos it is
difficult to produce a thin and long stream. However, for the halo
shapes that we find to be consistent with the Pal 5 and GD--1 data
below, this is not a major problem.

When we model the stream as being a single orbital torus, we can
further restrict the model. The parameter $\sigma_v$ determines the
location of the track, because it scales the frequency offset between
the progenitor and the stream, and it sets the width and length of the
stream (in conjunction with $t_d$). We do not attempt to match the
observed width of the streams below and therefore $\sigma_v$ only
affects the location of the track. That said, the $\sigma_v$ that
produce a matching tracks below also give reasonable stream
widths. For Pal 5 we fix the known sky position and line-of-sight
velocity of the progenitor and we use the dispersion parameter
$\sigma_v$ as a way to match the stream--progenitor offset. For GD--1,
for which we do not know the progenitor position, $\sigma_v$ would be
degenerate with the phase--space location of the progenitor and we
therefore fix it to a reasonable value. Similarly, one of the six
phase--space coordinates of the GD--1 progenitor is unconstrained by
this modeling and we fix one of the sky location coordinates near one
end of the GD--1 stream. Because we can approximate the stream track
as having constant frequency, it also does not matter whether we model
GD--1 as being a leading or trailing arm. This choice excludes the
possibility that GD--1's progenitor or dissolved progenitor sits in
the middle of the observed part of the stream. This is unlikely
because in this case a clear kink of $\Delta \phi_2 \approx 0.5^\circ$
would be visible at the $\phi_1$ position of the progenitor, which is
not observed.

\begin{figure*}
  \includegraphics[width=\textwidth,clip=]{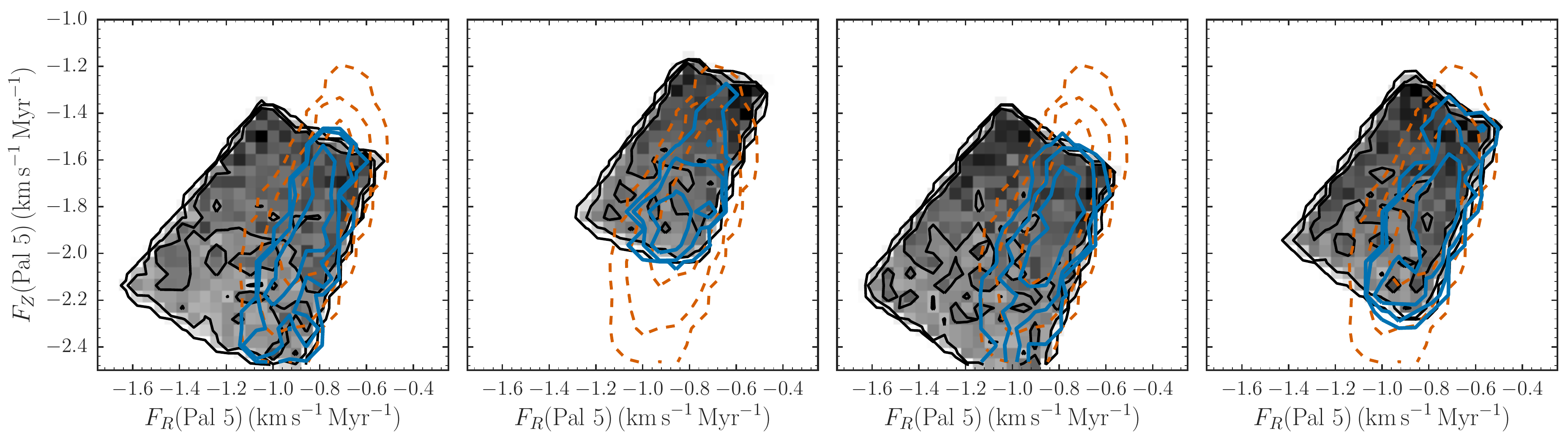}
  \caption{Example prior and posterior PDFs for the force at the
    fiducial location of the Pal 5 cluster for four of the 32
    considered potential families. These four examples were chosen to
    illustrate the range of force fields included in the modeling. The
    grayscale density and black contours display the prior on the
    force field (from varying the halo axis ratio and circular
    velocity uniformly within their allowed range for each potential
    family). The solid blue curves show the posterior PDF for each
    potential family. The dashed orange curve displays the posterior
    PDF considering all potential families simultaneously. Each
    potential family agrees on the constraint on the radial and
    vertical components of the force at the location of Pal 5 (to the
    extent allowed by the prior), even though their prior ranges are
    very different.}\label{fig:pal5postex}
\end{figure*}

The Milky-Way potential model of \sectionname~\ref{sec:mwmodel} is
characterized by 8 parameters and the stream model adds four (in the
case of Pal 5) or five (for GD--1) additional parameters. To lower the
dimensionality of each stream fit and thus speed it up, we restrict
the range of potentials as follows. We perform the fit described in
\sectionname~\ref{sec:priorfit} fixing $R_0 = 8\kpc$ and $V_c =
220\kms$ and we obtain 32 independent samples from the posterior PDF
for the remaining six parameters. For each of these 32 samples we then
analyze the stream data separately, keeping the five potential
parameters describing the relative amplitudes, the scale length and
height of the disk, and the scale radius of the halo fixed at the
value of the sample, but allowing $c/a$ and additionally $V_c(R_0)$ to
vary (between 1/2 and 2 for $c/a$ and between $200\kms$ and $250\kms$
for $V_c(R_0)$). These 32 sampled sets of fixed parameters are shown
as crosses in \figurename~\ref{fig:mwpot14_prior} versus the $c/a$ of
each sample, which is not held fixed. All of these are a priori
reasonable potential models, as $c/a$ and $V_c(R_0)$ are almost
uncorrelated with the other potential parameters (see
\figurename~\ref{fig:mwpot14_prior}). Thus, for each of the 32
samples, we sample a wide range of halo shapes and overall Milky-Way
masses and between the 32 families we sample a wide range of allowed
disk and halo models. Combining these 32 independent Markov Chain
Monte Carlo (MCMC) chains requires one in principle to compute the
relative evidence or marginal likelihood to weight them, but we find
that in each of the 32 potential families equally good fits to the
stream can be obtained and we simply weight all of the chains
equally. We will discuss this in more detail below, as it has
important implications for what information is obtained about the
gravitational potential from a single stream.

To transform between Galactocentric coordinates and heliocentric
coordinates, we fix $R_0 = 8\kpc$, the Sun's height above the
mid-plane $Z_0 = 25\pc$, and the Solar motion with respect to a
circular orbit to $(V_{R,\odot},V_{T,\odot},V_{Z,\odot}) =
(-11.1,24.0,7.25)\kms$ \citep{Schoenrich10a,Bovy15a}. Even a
$\approx10\kms$ uncertainty in $V_{T,\odot}$ only affects the inferred
force by $<10\,\%$, which is smaller than the statistical
uncertainties that we find below.

To summarize our stream-modeling methodology:\\$\bullet$ We only use
the observed location of the stream track, not the density along the
track or its width or extent.\\$\bullet$ We approximate each stream as
a single orbital torus, assuming a single frequency--offset between
the progenitor and every location along the observed part of the
stream. We fix the time at which disruption started to $t_d =10\Gyr$,
because it does not affect the location of the track in the
single-torus approximation and cannot be determined without an
observation of the extent of the stream;\\$\bullet$ We vary the
unknown or marginally-constrained phase--space coordinates of the
progenitor of the stream (the distance and proper-motion vector for
Pal 5 and the sky location, distance, proper motion, and line-of-sight
velocity for GD--1). For Pal 5 we vary the velocity-dispersion
parameter $\sigma_v$, but we do not for GD--1, because it is
degenerate with the unknown location of the progenitor.\\$\bullet$ We
consider a wide range of three-component potential models by sampling
32 plausible values of the relative disk--halo--bulge contribution and
of the scale parameters of the disk and halo and by for each of these
32 families fully varying the halo axis ratio $c/a$ and the circular
velocity $V_c(R_0)$ without any prior constraint.

\section{Analysis of Pal 5}\label{sec:pal5}

\subsection{Data and parameter sensitivity}

For Pal 5, the available data on the stream that we use are (a) the
stream sky position measurements from SDSS from
\citet{Fritz15a}\footnote{We note that there is a typo in Table 1 of
  \citet{Fritz15a}. The point at (RA,Dec) = (229.11,0.54) should
  instead be (229.61,0.54). The arXiv version of the paper has been
  corrected.} and (b) the line-of-sight velocity measurements of
individual stream members from \citet{Kuzma15a}. Because the latter do
not resolve the velocity dispersion within the stream, we use the
individual-member measurements as if they were measurements of the
line-of-sight velocity of the mean stream track. We use the position
measurements as the measured declination (Dec) of the track at a given
right ascension (RA) plus its uncertainty and the line-of-sight
velocities similarly as a function of RA. These data are shown in
\figurename~\ref{fig:pal5track}. While we will show the sky location
of the leading arm, we do not use it in the fit because it is so
short, but takes about as long to predict as the trailing arm.

\begin{figure*}
  \includegraphics[width=0.98\textwidth,clip=]{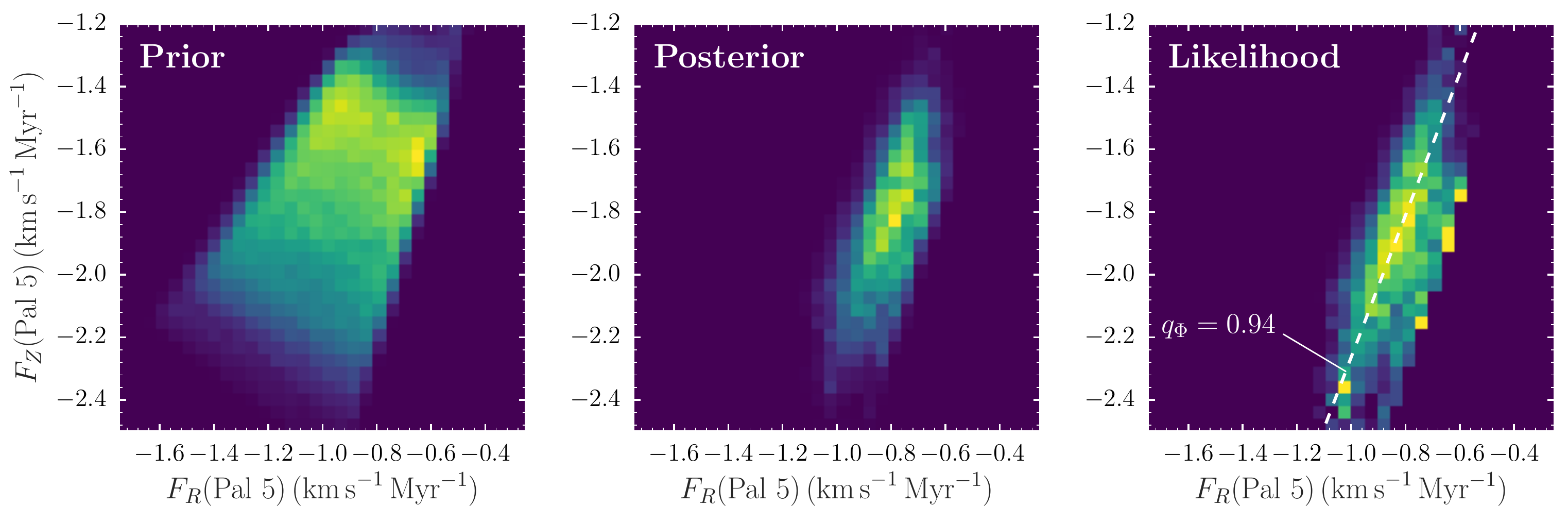}
  \caption{Constraint on the radial and vertical force components at
    the location of the Pal 5 cluster from the Pal 5 stream data. The
    left panel displays the prior on the force from all 32 potential
    families that we consider and the middle panel shows the posterior
    PDF combining all of the 32 independent potential-family MCMC
    chains. The right panel gives the ratio of the posterior and prior
    to demonstrate the measurement of the force coming from the Pal 5
    stream data. The likelihood is narrowest perpendicular to the
    direction of constant overall potential flattening: The dashed
    curve shows the location of constant $q_\Phi = 0.94$. A weaker
    constraint is obtained along the direction of constant
    flattening.}\label{fig:pal5post}
\end{figure*}

We further fix the following coordinates of the Pal 5 cluster as the
stream progenitor: RA = 229\degp 018, Dec = $-0$\degp124, and
$V_\mathrm{los} = −58.7\kms$. While the line-of-sight velocity has a
$\approx1\kms$ uncertainty, this is much smaller than the uncertainty
in the distance and proper motion of the Pal 5 cluster and it
therefore does not affect the stream track significantly. We use the
measurement of the proper motion of the Pal 5 cluster from
\citet{Fritz15a}: $(\mu_{\alpha}\,\cos \delta,\mu_\delta) =
(-2.296,-2.257) \pm (0.186,0.181)\mas\yr\inv$. However, the proper
motion uncertainty is large enough that it has a big impact on the
location of the stream track. We include the proper motion in our fits
and add the \citet{Fritz15a} measurement to the likelihood. Because we
find that the direction of the \citet{Fritz15a} proper-motion
measurement is almost exactly aligned with the direction of the
stream, we use an alternative parameterization of the proper motion in
terms of a component $\Delta \mu_\parallel$ in the direction of
$(2.296,2.257)$ and a component $\Delta \mu_\perp$ perpendicular to
this direction (toward $(2.257,-2.296)$). Measurements of the distance
to the Pal 5 cluster range from $20.9\kpc$ to $23.2\kpc$
\citep{Harris96a,Vivas06a,Dotter11a} and we conservatively assume a
flat prior on the distance between $20\kpc$ and $24\kpc$.

To get a sense of how the various model parameters affect the location
of the Pal 5 stream, we have computed the stream track varying the
halo axis ratio, the circular velocity, and the distance and parallel
proper-motion offset. As the fiducial model around which these
parameters are varied we assume a distance of $23.2\kpc$, the
\citet{Fritz15a} best-fit proper motion, a stream velocity-dispersion
parameters $\sigma_v = 0.2\kms$, and the best-fit three-component
potential to the data in \sectionname~\ref{sec:priorfit} obtained by
fixing $c/a=1$, $V_c(R_0) = 220\kms$, and $R_0 = 8\kpc$ (this
potential is almost exactly the same as \texttt{MWPotential2014},
except that its parameters have not been rounded to convenient values;
see \citealt{Bovy15a}). These track variations are displayed in
\figurename~\ref{fig:pal5track}. It is clear that the track, both on
the sky and in line-of-sight velocity, varies significantly when the
parameters are varied over a reasonable range. In particular, the
circular-velocity, the distance, and the proper motion of the
progenitor affect the stream track in a similar manner in both sky
position and velocity. Therefore, in fitting these data a model has
significant leeway to trade differences in $V_c(R_0)$ for differences
in the distance and proper motion of the progenitor.

However, it is also immediately clear from
\figurename~\ref{fig:pal5track} that changing the halo axis ratio,
which changes the overall flattening of the potential and thus the
ratio of the vertical and radial components of the force, changes the
sky position and the line-of-sight velocity of the stream track in a
manner that is perpendicular to changes induced by the other model
parameters. That is, when $V_c$ is increased, the distance is
decreased, or the proper motion is increased, the trailing tail of the
stream moves toward higher declinations and lower velocities. When the
halo axis ratio is increased, the trailing track moves toward both
higher declinations and higher velocities. Thus, when fitting the
data, the halo axis ratio can be measured independently from the other
parameters, which among themselves will have a high degree of
degeneracy.

\subsection{Pal 5 Potential constraints}

\begin{figure*}
  \begin{center}
  \includegraphics[width=0.9\textwidth,clip=]{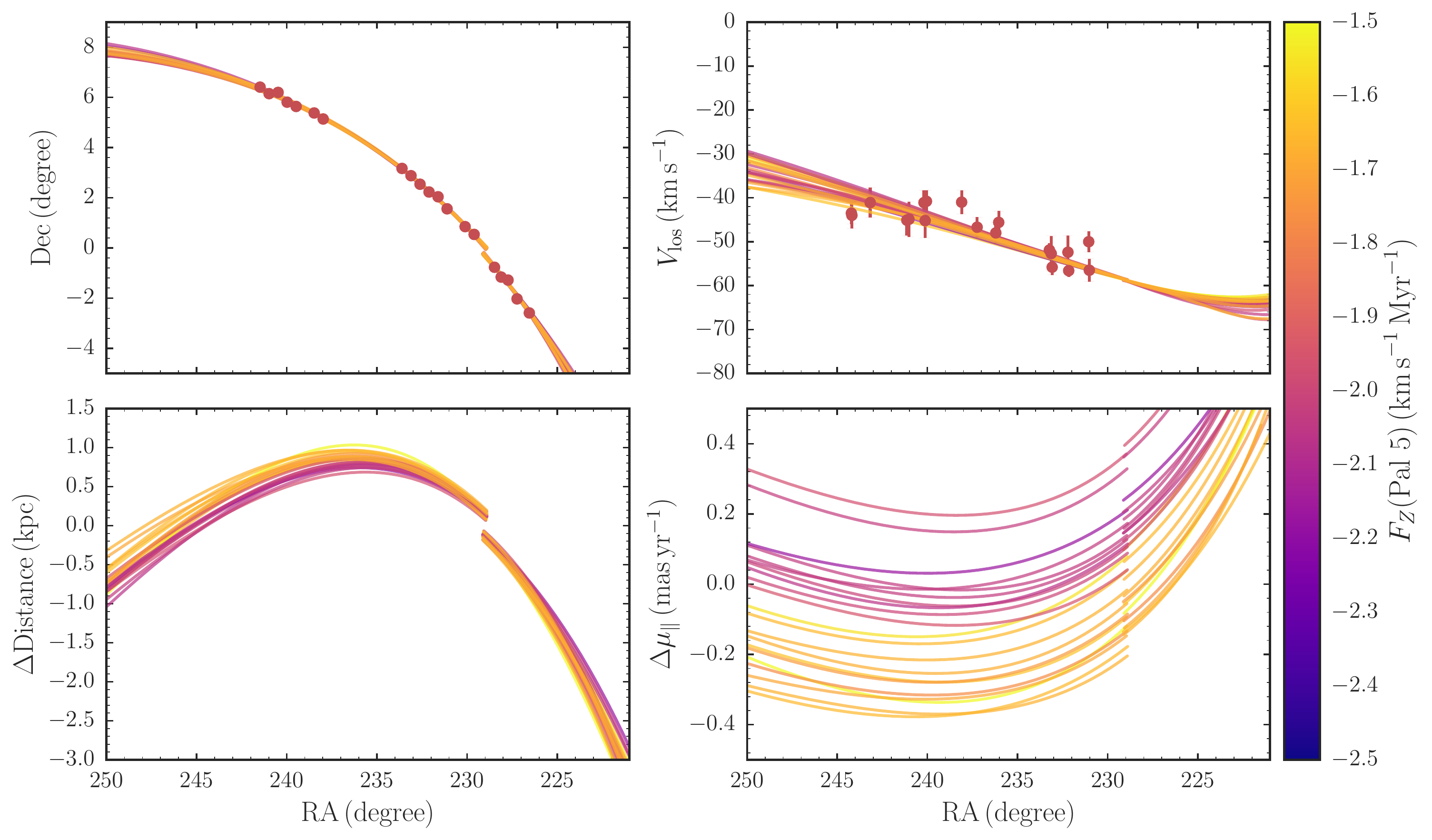}
  \end{center}
  \caption{Posterior samples from the fit to the Pal 5 stream
    data. The samples are color-coded by their value of the vertical
    component of the force at Pal 5 to illustrate how the measurements
    of the individual vertical and radial components of the force can
    be improved by future data. All of the displayed models fit the
    data well; the leading arm's sky location was not included in the
    fit, but is successfully recovered. Measurements of the sky
    location or radial velocity of the trailing arm at higher RA do
    not distinguish well between different forces. All models have
    approximately the same distance gradient along the stream. Better
    measurements of the proper motion of the Pal 5 cluster or stream
    would allow much improved measurements of the force field near Pal
    5.}\label{fig:pal5samples}
\end{figure*}

We explore the posterior PDF of each of the 32 potential families (see
above) using MCMC and combine all chains to provide a sampling of the
model parameters. We find the halo axis ratio to be $c/a =
0.9\pm0.2$. As expected, we find that the halo axis ratio is
uncorrelated with $V_c$ or the distance and parallel proper-motion
offset of the progenitor. There is a slight degeneracy between $c/a$
and the component of the proper motion perpendicular to the
\citet{Fritz15a} measurement, but the perpendicular component itself
is found to be well constrained ($\Delta \mu_\perp =
0.04\pm0.03\mas\yr\inv$) so this only has a small impact and a much
better proper-motion measurement would be necessary to reduce the
degeneracy. That our constraint on $\Delta \mu_\perp$ is so close to
zero is surprising given the $\approx0.2\mas\yr\inv$ uncertainty in
its direct measurement by \citet{Fritz15a}, but this is simply due to
chance. The parameters $V_c$, the distance $D_\mathrm{Pal\,5}$, and
parallel proper-motion offset are mutually degenerate: we find
$V_c(R_0) = 225\pm14\kms$, $D_\mathrm{Pal\,5} =
22.9^{+0.6}_{-0.9}\kpc$ (the asymmetric uncertainty is due to our
prior that $D_\mathrm{Pal\,5} < 24\kpc$), and $\Delta \mu_\parallel =
0.10\pm0.16\mas\yr\inv$. We also find that $\sigma_v \approx 0.2\kms$,
with only a minor correlation with $\Delta \mu_\perp$. This $\sigma_v$
leads to a stream width of $\approx20'$ FWHM, in good agreement with
the measurement from \citet{Carlberg12a}, especially considering that
we assume an age of $10\Gyr$ for the Pal 5 stream. This $\sigma_v$
also gives a line-of-sight velocity dispersion in the stream of
$\approx1\kms$, at the low end of, but consistent with the
measurements from \citet{Odenkirchen09a} and \citet{Kuzma15a}, which
give a velocity dispersion of $\approx2\kms$.

As already mentioned in \sectionname~\ref{sec:method} above, we find
that each of the 32 potential families essentially fits the stream
equally well (a single potential did not converge in a short enough
time and is not included in the following analysis, although it does
not change the results). These potentials explore a wide range of
disk-to-halo ratios, disk scale lengths, and halo
concentrations. Therefore, it is clear that the observed track of the
stream does not hold much information about the global properties of
the gravitational potential. An arm of a stream is approximately at a
constant frequency and therefore corresponds to a single orbital torus
along which energy should be conserved. This energy conservation is
ensured by balancing the work done by the kinetic energy when going
from one end of the stream to the other. Thus, we expect the force
along the stream to be a robustly measured quantity. To test this
hypothesis, we compute the force in all of the potentials that we
consider at a fiducial location for the Pal 5 cluster. That is, we fix
the distance to be $23.46\kpc$, which places the cluster at $(R,Z) =
(8.4,16.8)\kpc$; this conveniently has $Z/R = 2$.

Examples of prior and posterior PDFs for the radial and vertical
components of the force at the fiducial Pal 5 location are displayed
in \figurename~\ref{fig:pal5postex} for four of the 32 potential
families that we consider (chosen to illustrate the range of force
fields included in our models). The grayscale density shows the prior
on the radial and vertical components of the force when varying the
parameters of the potential model, $c/a$ and $V_c$ for each potential:
curves of constant $c/a$ run from the upper-right to the lower-left
edge, curves of constant $V_c$ run from the upper-left to the
lower-right edge. The blue contours display the posterior PDF for the
particular potential and the orange contours are the posterior PDF
from considering all of the potentials. This figure demonstrates that
even though the prior on the force can vary significantly among the 32
potential families, the posterior PDF always prefers the same force
(as much as allowed by each particular potential's prior on the
force).

The prior, posterior, and their ratio---the likelihood---considering
all potentials is shown in \figurename~\ref{fig:pal5post}. The prior
is obtained by sampling $c/a$ and $V_c$ from their prior range for all
32 potential families and computing the radial and vertical components
of the force for all of these potentials. The middle panel displays
the posterior PDF, which is much narrower than the prior. Because even
the uninformative, flat priors on $c/a$ and $V_c$ lead to a non-flat
prior for the force (the left panel), we need to divide the posterior
by the prior in the two-dimensional plane of the force components to
know what information about the force is obtained from the Pal 5 data
alone. This is shown in the left panel. The likelihood is similar to
the posterior, but slightly wider. This anisotropy in the likelihood
is expected from \figurename~\ref{fig:pal5track}: the flattening is
uncorrelated with the other model parameters, while the normalization
of the force is correlated with the distance and proper motion of the
Pal 5 cluster.

\begin{figure*}
  \includegraphics[width=\textwidth,clip=]{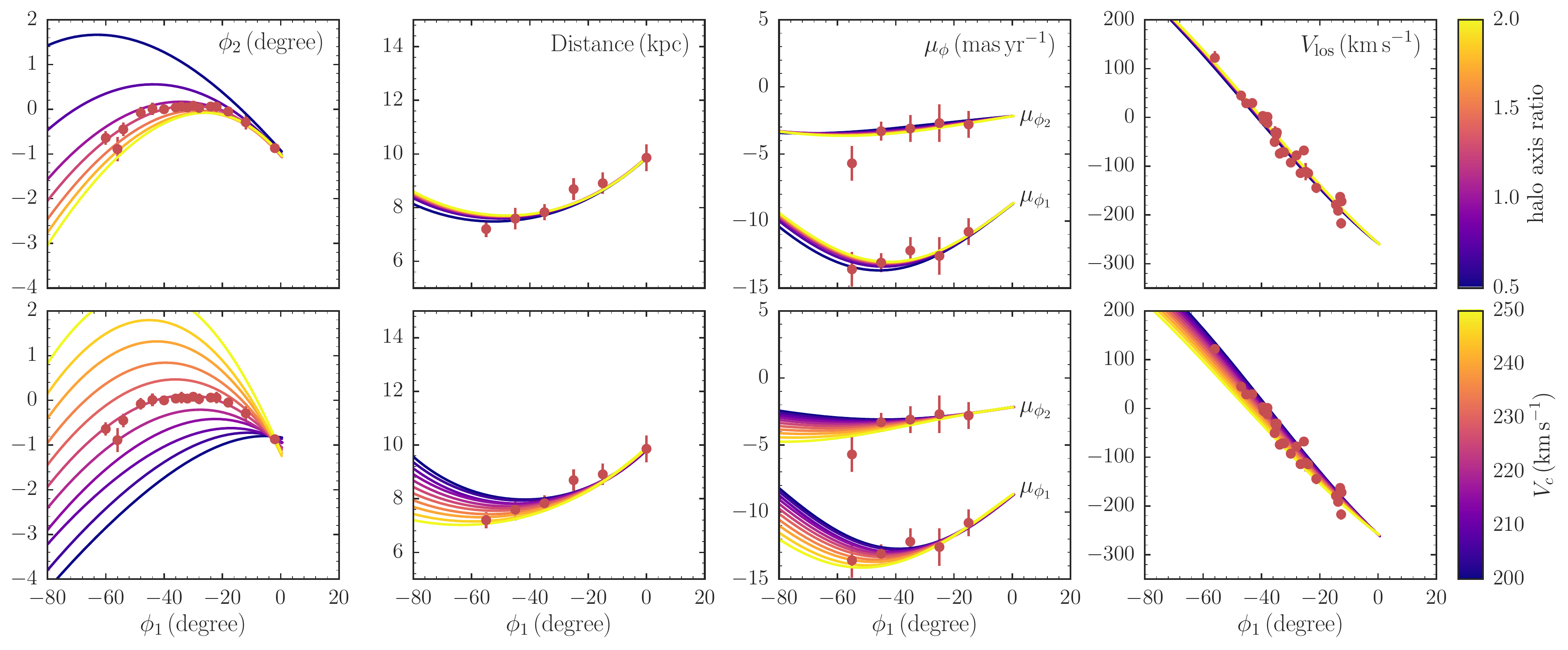}
  \caption{Sensitivity of the GD--1 stream track to variations in the
    gravitational potential. The top panels demonstrate the effect of
    changing the halo axis ratio, which effectively changes the ratio
    of the vertical to the radial components of the force near the
    GD--1 stream. The bottom panels show the effect of varying
    $V_c(R_0)$, which changes the magnitudes of the total force, while
    keeping their ratio the same. Only the stream's location on the
    sky (left panels) is sensitive to the halo axis ratio $c/a$; the
    distance or velocity components are hardly affected by changing
    $c/a$. All phase--space coordinates of the stream track are
    sensitive to $V_c$. Thus, the distance and velocity measurements
    constrain the location of GD--1's progenitor and $V_c$, while the
    sky location allows the halo axis ratio to be
    measured.}\label{fig:gd1track}
\end{figure*}

The likelihood in \figurename~\ref{fig:pal5post} displays a clear
degeneracy between the radial and vertical components of the
force. The dashed line shows the location of a constant flattening of
the total potential, defined as
\begin{equation}\label{eq:flat}
  q^2_\Phi = \frac{Z}{R}\,\frac{F_R}{F_Z}\,,
\end{equation}
where $Z/R = 2$ for Pal 5. It is clear that the degeneracy between
$F_R$ and $F_Z$ runs along the line of constant $q_\Phi$. The
constraint in the direction perpendicular to this line is three times
narrower than that parallel to the line.

We can therefore summarize the information about the Milky Way's
gravitational potential learned from the Pal 5 stream data in the
direction of the two principal directions of the likelihood in
\figurename~\ref{fig:pal5post} in the following convenient manner:
\begin{align}\label{eq:pal5qm}
q_\Phi & = 0.94\pm0.05\,
\end{align}
and
\begin{align}\label{eq:pal5fm}
0.94^2\,(F_R+0.80)+2\,(F_Z+1.82) & = -0.2\pm0.6
\end{align}
where $F_R$ and $F_Z$ are measured in $\kms\Myr\inv$ at $(R,Z) =
(8.4,16.8)\kpc$ and $\phi = 178$\degp 4.

Samples from the MCMC chains are compared with the data in
\figurename~\ref{fig:pal5samples}. All of these fit the data well. As
discussed above, we did not include the sky location of the leading
arm in the fit, but it is nevertheless well recovered. We also show
the predicted track of the stream in distance and (parallel) proper
motion. The weak distance gradient is consistent with the weak
gradient observed by \citet{Ibata16a}. The samples are color-coded by
the value of the vertical force component at Pal 5's location, to
illustrate how the measurement in \equationname~\eqref{eq:pal5fm}
could be improved by future data. It is clear that better measurements
of the sky location or line-of-sight velocity, even at larger
distances from the cluster, do not distinguish between different
values of $F_Z$ well. Similarly, measurements of the distance gradient
along the stream will not improve the force measurements, because the
gradient is similar in all allowed models. However, a better
measurement of the proper motion of the Pal 5 cluster or a measurement
of the stream's proper motion (both with $\sigma_\mu \lesssim
0.1\mas\yr\inv$) would significantly improve the force measurements.

\subsection{Comparison to K\"{u}pper et al.}

A previous analysis of similar Pal 5 stream data was performed by
\citet{Kuepper15a}, who claim tight constraints on the mass, scale
radius, and potential-flattening $q^h_\Phi$ of the dark--matter halo
based on these data. We find, however, that the Pal 5 stream data on
their own contain very little information about the mass and scale
radius of the dark--matter halo. To investigate why our results differ
from those of \citet{Kuepper15a}, we perform some simple
experiments. First, we sample the parameters of the potential model
used by \citet{Kuepper15a} using the same flat priors as
\citet{Kuepper15a}, but requiring that $V_c(R_0)$ is between $200\kms$
and $280\kms$ (their prior on $V_c(R_0)$) and that the overall
potential flattening is constrained to be $q_\Phi = 0.94\pm0.05$, as
we find from the Pal 5 data. We find no preference for any mass or
scale radius of the dark--matter halo in this case, except for a
strong degeneracy between these quantities because of the limited
range in $V_c$. However, we find that $q^h_\Phi = 0.95\pm0.15$ in
good agreement with \citet{Kuepper15a}. We can add in a constraint
similar to that found in \equationname~\eqref{eq:pal5fm}, but changed
to be consistent with \citet{Kuepper15a}'s best-fit total force
amplitude (which is consistent with ours within the uncertainties). In
this case, we still find that there is hardly any constraint on the
mass or scale radius of the halo, except that the allowed mass range
for a given scale radius is reduced; the constraint on $q^h_\Phi$
remains the same.

\begin{figure*}
  \includegraphics[width=0.98\textwidth,clip=]{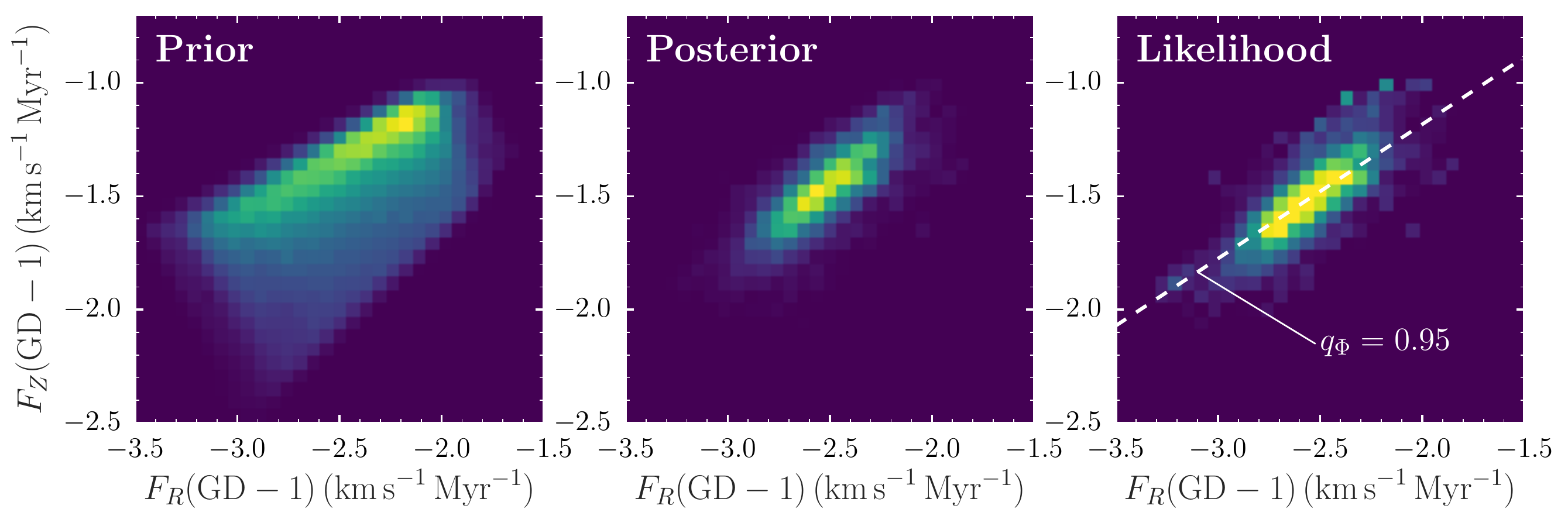}
  \caption{Like \figurename~\ref{fig:pal5post}, but for the GD--1
    analysis. As for Pal 5, the likelihood is narrowest perpendicular
    to the direction of constant overall potential flattening. The
    dashed curve displays the location of $q_\Phi = 0.95$, the
    best-fit potential flattening. The GD--1 data also provide a good
    constraint on the component of the force along the direction of
    constant flattening.}\label{fig:gd1post}
\end{figure*}

Thus, we conclude that our measurements in \equationname
s~(\ref{eq:pal5qm}) and (\ref{eq:pal5fm}) are weaker than those
claimed by \citet{Kuepper15a}. This difference may be due to the fact
that \citet{Kuepper15a} ascribe and fit overdensities along the stream
as resulting from pericentric stripping episodes, thus in essence
adding a constraint on the radial frequency of the orbit. Because the
significance of the observed overdensities is contested
\citep{Thomas16a} and because overdensities could be due to subhalo
encounters \citep[\eg,][]{Carlberg12a,Bovy16a}, not employing the
overdensities as a constraint on the gravitational potential appears
to be the conservative thing to do. In summary, our measurements of
$q_\Phi$ and the force at Pal 5 are consistent with, but more
conservative than the measurements of \citet{Kuepper15a}.

\section{Analysis of GD--1}\label{sec:gd1}

\subsection{Data and parameter sensitivity}

For GD--1, we fit the data from \citet{Koposov10a}, which is shown in
\figurename~\ref{fig:gd1track}. These data come in the form of the
location of the track in a set of custom sky coordinates
$(\phi_1,\phi_2)$ with $\phi_1$ approximately aligned with the
stream. We use the measurements of the stream phase--space location as
a function of $\phi_1$ in $\phi_2$, distance, proper motion, and
line-of-sight velocity. Similar to the analysis of the Pal 5 data
above, we also treat the line-of-sight velocity measurements of stream
members as measurements of the track location in line-of-sight
velocity at each star's $\phi_1$, because these measurements do not
resolve the velocity dispersion of the stream. Because for GD--1 we
have data on all six phase--space coordinates of the stream itself,
the location of the progenitor---in our approximation of the stream
track---is much better constrained than that of Pal 5 (especially
because the distance to GD--1 is much better constrained than that to
Pal 5 above). As discussed above, we fix the progenitor's location to
$\phi_1 = 0^\circ$ and $\sigma_v = 0.4\kms$ and thus model the GD--1
stream as being a leading arm. We stress that this is simply the
location of the progenitor in our approximation, not the true location
of the progenitor.

We explore the sensitivity of the stream track to the gravitational
potential in \figurename~\ref{fig:gd1track}. Similar to the
sensitivity analysis for Pal 5 in \figurename~\ref{fig:pal5track}, we
use the bulge, disk, and halo parameters of the best-fit
three-component potential to the data in
\sectionname~\ref{sec:priorfit} obtained by fixing $c/a=1$, $V_c(R_0)
= 220\kms$, and $R_0 = 8\kpc$ as the fiducial potential family, but we
find the best-fitting $c/a$, $V_c(R_0)$, and GD--1 progenitor
parameters by fitting to the GD--1 data in order to have a fiducial
model that fits the data well for this sensitivity analysis. We then
vary $c/a$ and $V_c$ around their best-fitting values and compute the
stream track in all phase--space components. The results from this are
displayed in \figurename~\ref{fig:gd1track}.

It is clear from \figurename~\ref{fig:gd1track} that varying the halo
axis ratio has almost no effect on the stream's track in distance or
any of the velocity components. Changing $V_c$ has a small effect in
the same components. However, the stream's location on the sky varies
significantly for different $c/a$ or $V_c$. Within this potential
family, $V_c$ can be constrained by the stream's distance, proper
motion, and line-of-sight velocity independent of $c/a$ and the
stream's location in the sky can then essentially constrain
$c/a$. Thus, we again expect little to no correlation between the
constraints on $c/a$ and those on the other parameters of the model
for the GD--1 data. $V_c$ will similarly again be somewhat degenerate
with the phase--space location of the progenitor.

\subsection{GD--1 Potential constraints}\label{sec:gd1constraints}

We explore the posterior PDF for the potential and progenitor
parameters for GD--1 using MCMC for the 32 potential families. Because
for GD--1, unlike for Pal 5, we have measurements of all six
phase--space components the fit is more strongly constrained and every
potential family prefers similar parameters. We find that $c/a =
1.27^{+0.27}_{-0.22}$ with no significant correlations with any of the
other model parameters. The circular velocity is constrained to be
$V_c(R_0) = 225\pm10\kms$, correlated with the proper motion of the
progenitor. The progenitor is at
$(\phi_1,\phi_2,D,\mu_{\phi_1},\mu_{\phi_2},V_{\mathrm{los}}) =
(0^\circ,-0$\degp82$\pm0$\degp
08$,10.1\pm0.2\kpc,0.0\pm0.3\mas\yr\inv,-0.15\pm0.10\mas\yr\inv,-257\pm5\kms)$. The
only significant correlations between the progenitor parameters are
between the proper motion components.

Similar to the fit to Pal 5 above, we find that each of the 32
potential families provides an equally-good fit to the GD--1
data. Thus, the GD--1 data again do not hold much information about
the global properties of the gravitational potential. A similar
investigation to that shown in \figurename~\ref{fig:pal5postex}
demonstrates again that while the effective prior on the local radial
and vertical force components differs significantly between the 32
potential families, the posterior PDF of the radial and vertical force
components near GD--1 is almost exactly the same in all 32
cases. Here, we compute the force at $(R,Z) = (12.5,6.675)\kpc$ and
$\phi=0^\circ$, chosen to be close to the center of the GD--1 stream
data.

\begin{figure*}
  \begin{center}
  \includegraphics[width=0.9\textwidth,clip=]{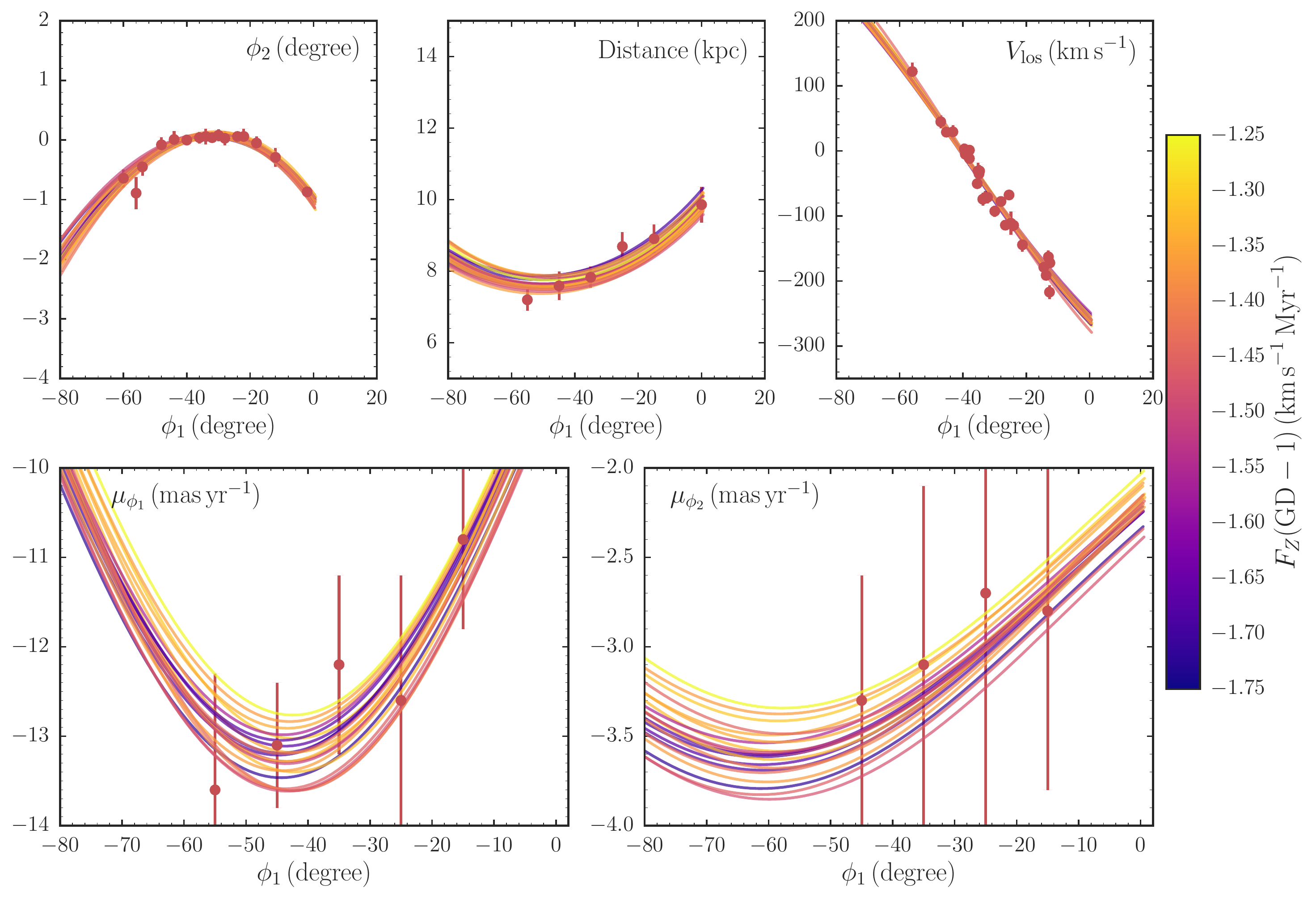}
  \end{center}
  \caption{Posterior samples from the GD--1 stream fit. As in
    \figurename~\ref{fig:pal5samples}, the samples are color-coded by
    their value of the vertical force component at GD--1 to illustrate
    how the measurements of the individual vertical and radial
    components of the force can be improved by future data. All of the
    displayed models fit the six-dimensional phase--space data for
    GD--1 well. Better measurements of the sky location, distance, or
    line-of-sight do not distinguish well between different
    forces. Improved, sub $\mas\yr\inv$ measurements of the proper
    motion along the GD--1 stream would allow much improved
    measurements of the force field near GD--1.}\label{fig:gd1samples}
\end{figure*}

The prior and posterior PDF for the radial and vertical force
components at the center of the GD--1 data is displayed in
\figurename~\ref{fig:gd1post} combining the results from all 32
potential families. It is clear that the posterior is much more
sharply peaked than the prior and that the GD--1 data therefore
strongly constrain the force field in their vicinity. The ratio of the
posterior and the prior (the likelihood) is shown in the right
panel. Much as for Pal 5, the likelihood is anisotropic with the
principal axes approximately aligned with the directions of constant
overall potential flattening $q_\Phi$ and the direction perpendicular
to this. For GD--1, the likelihood peaks at $q_\Phi = 0.95$. The ratio
of the widths in the two principal directions is again about three.

We can thus summarize the information about the gravitational
potential provided by the GD--1 data alone as
\begin{align}\label{eq:gd1qm}
q_\Phi & = 0.95\pm0.04\,
\end{align}
and
\begin{align}\label{eq:gd1fm}
0.95^2\,(F_R+2.51)+6.675/12.5\,(F_Z+1.47) & = -0.05\pm0.3
\end{align}
where $F_R$ and $F_Z$ are measured in $\kms\Myr\inv$ at $(R,Z) =
(12.5,6.675)\kpc$ and $\phi = 0^\circ$.

Samples from the MCMC chains for GD--1 are compared with the data in
\figurename~\ref{fig:gd1samples}. All of these match the data
well. The samples are color-coded by their value of the vertical force
component near the center of the stream, to illustrate how the
measurement in \equationname~\eqref{eq:gd1fm} could be improved by
future data. It is clear that the main improvement would again come
from sub $\mas\yr\inv$ measurements of the proper motion along the
stream.

\subsection{Comparison to Koposov et al. and Bowden et al.}

\begin{figure*}
  \includegraphics[width=\textwidth,clip=]{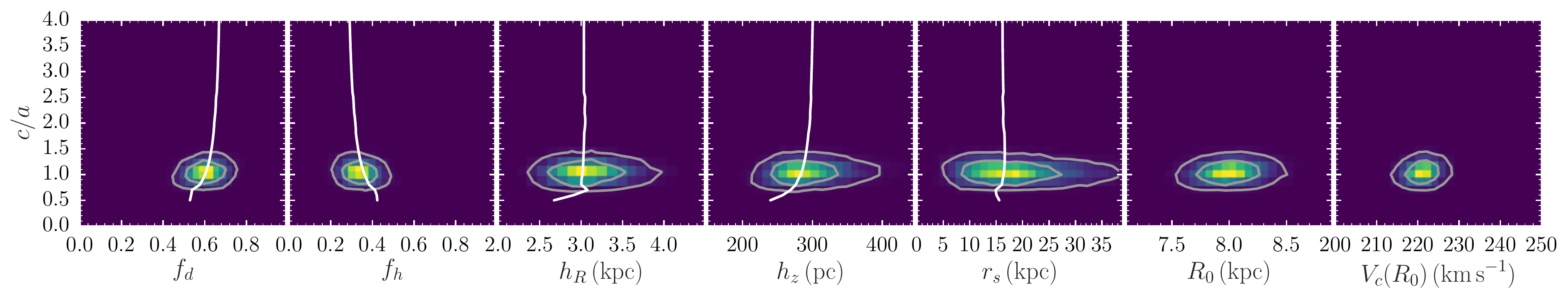}
  \caption{Constraints on the halo-axis ratio $c/a$ and the disk,
    bulge, and halo components of the Milky Way after adding the
    constraints derived from the Pal 5 and GD--1 stream to the data in
    \sectionname~\ref{sec:priorfit}. Compared to the weak constraint
    on $c/a$ from prior dynamical data in
    \figurename~\ref{fig:mwpot14_prior}, the Pal 5 and GD--1 data
    provide a good measurement of $c/a = 1.05\pm0.14$ with no
    correlation with the other properties of the
    potential.}\label{fig:mwpot14_post}
\end{figure*}

The GD--1 data that we fit in the previous subsection were previously
analyzed by \citet{Koposov10a} and \citet{Bowden15a}. Because
\citet{Koposov10a} fit an orbit to these data, which is improper and
has been superseded by the analysis of \citet{Bowden15a}, we focus on
the latter, but briefly return to \citet{Koposov10a} at the end of
this subsection. \citet{Bowden15a} perform a similar fit as the one in
this paper, fitting a stream track to the phase--space location of the
data, but they compute the model stream tracks using orbit integration
of tracer particles stripped from a model progenitor and they assume a
simple flattened logarithmic potential rather than the three-component
potential that we use. Apart from the difference in potential, this is
essentially the same approach as we follow here, except that their
prescription for the phase--space distribution of the tidal debris at
the time of stripping is different from ours in detail and that we use
action--angle coordinates to compute a smooth representation of the
present-day track, while they use Monte-Carlo orbit integration to
reconstruct the present-day track. Thus, we expect to find essentially
the same results as \citet{Bowden15a}. However, our constraints on the
potential flattening and $V_c$ are significantly tighter than those
found by \citet{Bowden15a}.

To investigate the cause of this difference, we have repeated our fit
to the GD--1 data, but using a flattened logarithmic potential with
free parameters $q_\Phi$ and $V_c$ ($\Phi =
V_c^2/2\ln[R^2+z^2/q_\Phi^2]$), setting $R_0 = 8.5\kpc$ (but we get
similar results for $R_0=8\kpc$), and using the same flat priors as
\citet{Bowden15a}. In this case the fit is much faster than for our
three-component potential families, because orbit integration in this
simple potential is much faster, and we can obtain 10,000 MCMC samples
in a matter of hours. We find that $q_\Phi = 0.90\pm0.06$ and $V_c =
212\pm15\kms$ in this case. The constraint on $q_\Phi$ is weaker than
in the case of the three-component potential models and is in good
agreement with that of \citet{Bowden15a}; we also recover the same
correlation between $q_\Phi$ and $V_c$. Our best-fit value for $V_c$
is somewhat smaller than that of \citet{Bowden15a} although within the
formal uncertainties, which may be due in part because of our higher
assumed solar-motion component $V_{T,\odot}$.; the uncertainty in
$V_c$ is similar. Overall, the two analyses agree well.

Therefore, we conclude that our modeling of the GD--1 data in terms of
three-component potential families provides tighter constraints on the
gravitational potential than using a simple logarithmic
potential. This is probably due, at least in part, to the fact that
the simple logarithmic potential does not realistically describe the
structure of the potential. As shown by \citet{Williams14a}, the
offset between a tidal stream and a single orbit is very small for the
flattened logarithmic potential and \citet{Bowden15a} also find that
simply fitting the GD--1 data as an orbit returns the same results as
their stream fit. However, GD--1 is close enough to the disk that in a
more realistic potential the stream--orbit offset is larger and varies
with $q_\Phi$ \citep{Sanders13a}. Comparing orbits to the stream
tracks for Pal 5 and GD--1 shown in \figurename s~\ref{fig:pal5track}
and \ref{fig:gd1track} directly demonstrates that the location of the
tracks varies more than that of orbits when changing the
potential. Provided that one is willing to assume that the potential
is closer to a disk-plus-halo than to a logarithmic potential, this
effect should lead to stronger constraints on the potential and our
modeling of the potential in terms of three-component potential
families capitalizes on this.

Finally, \citet{Koposov10a} claim that the GD--1 data cannot constrain
the flattening of the halo without informative priors on the disk and
only find a lower limit $q_\Phi^h \gtrsim 0.9$ from the GD--1
data. We, however, find that the GD--1 data do provide an informative
constraint on the shape of the halo. This is in part due to the fact
that we only allow reasonable values for the mass of the disk. Even
though our potential families explore a wide range of possible disk
parameters, the disk always has $M_{\mathrm{disk}} \lesssim
10^{11}\msun$, while \citet{Koposov10a} allow unrealistic values above
this value and have a maximum-likelihood $M_{\mathrm{disk}} \approx
1.4\times10^{11}\msun$. However, this is not the only cause of our
better constraint, because we find the same best-fit value of $c/a
\approx 1.25$ for all potential families and furthermore, $c/a$ is
uncorrelated with $V_c(R_0)$, which scales the mass of the disk within
each family. Thus, the main improvement comes from properly modeling
the GD--1 data as a stream in a three-component potential, which leads
to tighter constraints on the potential and on $c/a$ in particular. It
is clear that the Pal 5 data provide a more direct measurement of the
shape of the halo than the GD--1 data, because the Pal 5 data are
located much further from the disk than the GD--1 data.

\section{Combined constraints on the shape of the inner dark--matter halo}\label{sec:combined}

In sections~\ref{sec:pal5} and \ref{sec:gd1} above, we inferred values
of the halo axis ratio of $c/a = 0.9\pm0.2$ and $c/a =
1.27^{+0.27}_{-0.22}$ for the Pal 5 and GD--1 streams. These
measurements were obtained by fitting general potential families,
inspired by fits to data on the rotation curve, vertical kinematics,
etc. from \sectionname~\ref{sec:priorfit}, but not required to provide
a good fit to those data. We saw that each of 32 potential families
provided an equally good fit to the stream data and that the inferred
value of $c/a$ was uncorrelated with $V_c$, indicating that these
$c/a$ values should not be significantly changed if we further require
the potential to fit the constraints from
\sectionname~\ref{sec:priorfit}. In this section we perform fits to
the data in \sectionname~\ref{sec:priorfit} plus the measurements for
Pal 5 from \equationname s~(\ref{eq:pal5qm}) and (\ref{eq:pal5fm}) and
for GD--1 from \equationname s~(\ref{eq:gd1qm}) and (\ref{eq:gd1fm}).

First, we add the Pal 5 and GD--1 constraints separately, obtaining
$c/a = 0.93\pm0.16$ for Pal 5 and $c/a = 1.3^{+0.5}_{-0.3}$ for GD--1,
in line with the measurements from the stream data directly, but with
slightly different uncertainties. The uncertainties for GD--1 are
somewhat larger than those found from considering the 32 potential
families in \sectionname~\ref{sec:gd1constraints} because we no longer
demand that $c/a < 2$---the prior in the analysis in
\sectionname~\ref{sec:gd1constraints}---and because we only use the
GD--1 constraints on the force components, not the full GD--1
data. The $c/a$ values obtained for Pal 5 and GD--1 are uncorrelated
with any of the other potential parameters describing the bulge, disk,
and halo or with $R_0$ and $V_c(R_0)$. Because GD--1 is located at a
Galactocentric radius of $\approx14\kpc$ and Pal 5 at a radius of
$\approx19\kpc$, this hints at a decrease in the axis ratio with
Galactocentric radius, although the values are entirely consistent
with each other.

Fitting all constraints simultaneously, we find that $c/a =
1.05\pm0.14$, again with no correlation with any of the other
parameters describing the potential. The $2\sigma$ range spans $c/a =
0.79$ to $c/a = 1.33$. \figurename~\ref{fig:mwpot14_post} displays the
joint two-dimensional posterior PDF for $c/a$ versus all of the other
model parameters, similar to \figurename~\ref{fig:mwpot14_prior}
before the addition of the stream data. The halo is thus constrained
to be spherical to about $15\,\%$. The scale radius of the
dark--matter halo is constrained to be $18.0\pm7.5\kpc$, which is a
significant improvement over the constraint without the Pal 5 or GD--1
data (which was $22^{+30}_{-9}\kpc$). This improvement is due to the
additional force measurements at $r\approx14\kpc$ and
$r\approx19\kpc$, which allow the radial profile of the halo to be
better determined. The mass of the dark--matter halo within 20 kpc is
also well constrained by these data: $M_{\mathrm{halo}}(r<20\kpc) =
1.1\pm0.1\times10^{11}\msun$ (we define the mass as the average of
$-r^2\,F_r$ at constant spherical $r=20\kpc$ for the non-spherical
dark--matter halos, but the exact definition is unimportant as we find
the halo to be consistent with spherical). Such a halo mass and scale
radius are fully consistent with the expected mass and concentration
for the Milky Way's halo.

To investigate how these results depend on the assumed density
distribution of the disk, we have repeated this fit for a
double-exponential disk model without and with an additional gas
component that is the same as that in
\sectionname~\ref{sec:priorfit}. The best-fit value of the halo axis
ratio in both cases is slightly, but not very significantly, lower:
$c/a = 0.99\pm0.14$ and $c/a = 0.98\pm0.13$ for without and with gas,
respectively; the $2\sigma$ lower limit is $c/a = 0.75$ in both
cases. The mass within 20 kpc and the scale radius of the halo are
almost unchanged. The detailed mass profile of the disk therefore only
has a minor effect. The scale length and height of the disk in all of
the disk models, or $R_0$ and $V_c(R_0)$ are as expected not
significantly better constrained after adding in the stream
constraints.

As discussed in the introduction of this paper, we expect the Milky
Way, with its massive, maximal disk \citep{Bovy13a} to have $c/a
\approx0.8$. This is almost $2\sigma$ below our best value for $c/a$
(the $2\sigma$ lower limit is $c/a = 0.79$). Thus, our finding that
the halo is spherical is in tension with the predictions from
numerical simulations. Better data on the Pal 5 and GD--1 streams
from, \eg, \emph{Gaia} or similar data on additional streams will
allow us in the near future to better determine whether this tension
is real or not.

To summarize current measurements of the force field in the inner
Milky Way, we display the force field in the best-fit model in
\figurename~\ref{fig:forcefield}. We include direct measurements of
the forces from the Pal 5 and GD--1 streams from this paper to
demonstrate how they agree with those derived from the joint fit. We
also include a summary of the results on the rotation curve
\citep{Bovy12a} and vertical-force curve \citep{Bovy13a}. Because
these are not measured at the same location (the rotation curve being
measured at $Z = 0$ and the vertical force measured at $|Z| =
1.1\kpc$), we move the radial-force measurements from the rotation
curve to $|Z| = 1.1\kpc$ using the fact that $\partial F_R / \partial
Z = \partial F_Z / \partial R$ \citep{Bovy12b} and assuming the $
\partial F_Z / \partial R$ value measured by \citet{Bovy13a}. While
there remains a large volume to be explored, this figure clearly
demonstrates the improvement in our understanding of the halo's
potential and its shape provided by the Pal 5 and GD--1 force
measurements.

\section{Conclusion}\label{sec:conclusion}

In this paper we have performed the first proper stream fits to
observational data using the action--angle approach to modeling
stellar streams \citep[\eg,][]{Eyre11a,Bovy14a,Sanders14a}. This is
also the first analysis to combine data from multiple streams to
improve our knowledge of the Milky Way's gravitational potential
within 20 kpc from the center.

Comparing our fits to previous analyses of the same (for GD--1) or
similar (for Pal 5) data, we note that our analysis is much faster
while simultaneously exploring a wider range of potentials than that
considered by \citet{Bowden15a} for GD--1 and \citet{Kuepper15a} for
Pal 5. Because the action--angle approach directly provides a smooth
stream track that can be computed using $\approx60$ orbit integrations
per model, we can evaluate the likelihood of different models in tens
of seconds. This is the case, even though we are modeling the
dark--matter halo using the flattened density of
\equationname~\eqref{eq:nfw}, for which computing the potential and
forces is non-trivial and requires numerical integration. By
approximating the stream as a single orbital torus over the
observable, high-surface brightness part of the stream, we also keep
the number of parameters the same as in the orbit fit of GD--1 by
\citet{Koposov10a} (for a single of our 32 potential families), which
means that our fit explores a volume of the same dimensionality as in
the simple orbit approximation, while properly fitting a stream. Thus,
we are able to fit the stream without resorting to high-performance
computing and are thus able to easily fit multiple streams.

By exploring a wide range of potential families for both the Pal 5 and
GD--1 streams, we have shown that the only information gained about
the Milky Way's gravitational potential appears to be the local
forces. This is because we can fit these streams equally well for all
a priori likely potentials that explore a wide range of disk-to-halo
ratios, disk profiles, and halo radial profiles. However, in each of
32 potential families that we consider for each stream, the a
posteriori probable forces in the vicinity of the stream are the
same. Despite this, we find that the \emph{form} of the potential does
matter in the fit, because comparing the results of our
three-component potential fits to a simple logarithmic-potential fit,
we find tighter and somewhat different constraints on the local
forces. This is due to the fact that our three-component potential
families properly take into account the effect of the varying offset
between the stream track and an orbit for different potentials, which
is absent in the less realistic logarithmic potential.

\begin{figure}
  \includegraphics[width=0.49\textwidth,clip=]{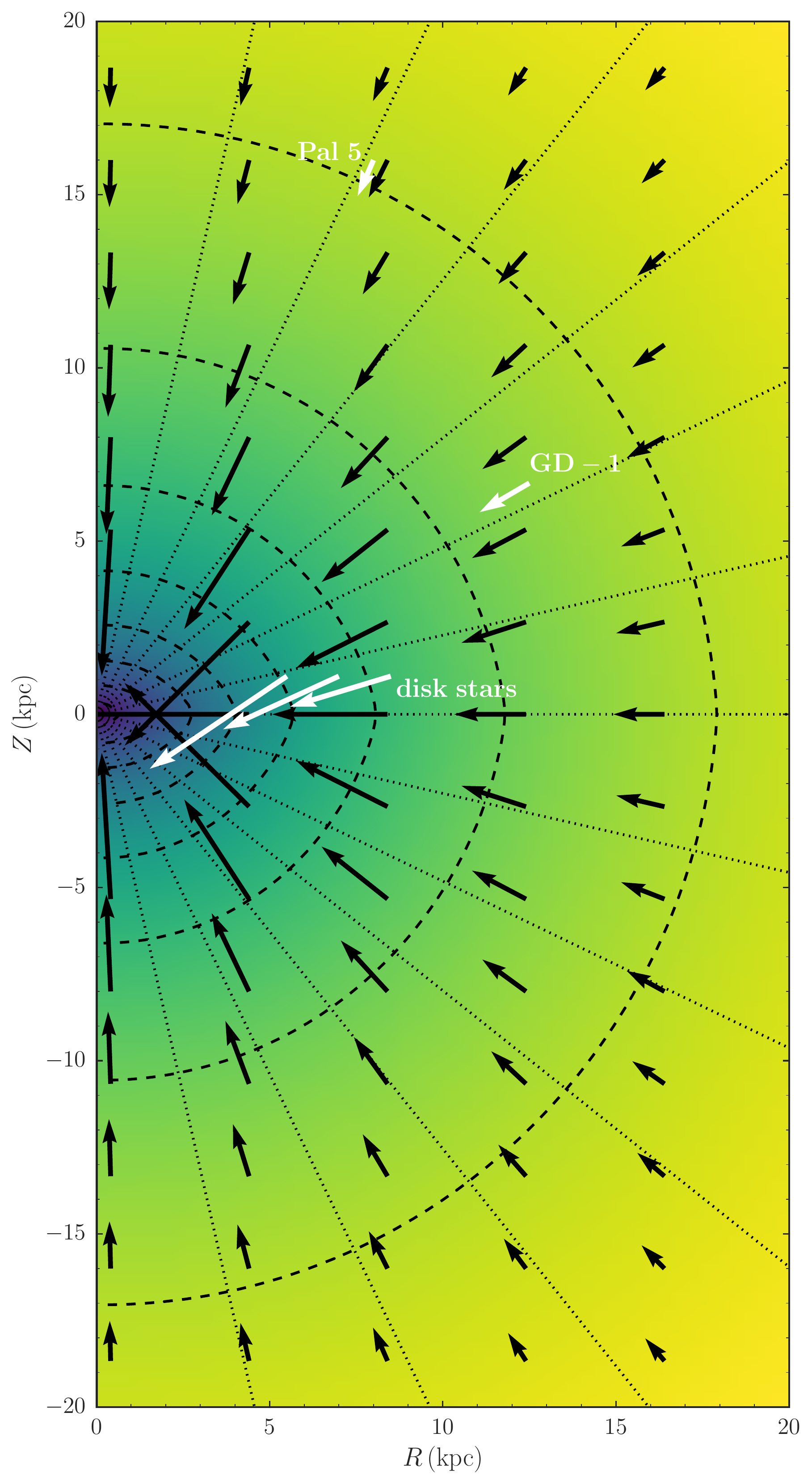}
  \caption{Best-fit force field for the Milky Way. The vectors display
    the two-dimensional radial and vertical components of the force
    field in the best-fit model to the data in
    \sectionname~\ref{sec:priorfit} and the data derived from the Pal
    5 and GD--1 streams in this paper. The color map and dashed
    contours show the value of the gravitational potential itself. The
    dotted lines go radially outward from the Galactic center to
    indicate the direction of a force pointing towards the center. The
    white vectors are direct measurements of the forces: the
    combination of the radial and vertical forces from \citet{Bovy12a}
    and \citet{Bovy13a} (labeled ``disk stars'' and summarized into
    three vectors), the force at the location of the Pal 5 cluster
    from \sectionname~\ref{sec:pal5}, and the force at the location of
    the GD--1 stream from
    \sectionname~\ref{sec:gd1}.}\label{fig:forcefield}
\end{figure}

For both the Pal 5 and GD--1 streams, we find that the ratio of the
vertical-to-radial components of the local force is more strongly
constrained than their overall amplitude. This is because, when
fitting the phase--space data for both streams, an overall
force-amplitude change can be traded off for differences in the
distance or proper motion of the stream or progenitor (see \figurename
s~\ref{fig:pal5track} and \ref{fig:gd1track}). Therefore, we express
our constraints on the gravitational potential from both streams in
terms of a measurement of the potential flattening $q_\Phi$ on the one
hand and in terms of a constraint on the force vector projected onto
the direction of constant flattening. These constraints are given in
\equationname s~\eqref{eq:pal5qm} and \eqref{eq:pal5fm} for Pal 5 and
\equationname s~\eqref{eq:gd1qm} and \eqref{eq:gd1fm} for GD--1. All
of these constraints have divided out the effective prior due to the
potential families used in the fit and are therefore pure measurements
based on the data for Pal 5 and GD--1 alone. The new force
measurements are summarized in \figurename~\ref{fig:forcefield}, which
demonstrates that they agree very well with the overall best-fit force
field to a variety of dynamical data in addition to the stream data
from this paper.

Because the streams considered here are exquisitely sensitive to the
flattening of the potential, they provide a strong constraint on the
axis ratio $c/a$ of the dark--matter halo's density. For GD--1 and Pal
5 individually we measure $c/a = 1.3^{+0.5}_{-0.3}$ for GD--1
($r\approx14\kpc$) and $c/a = 0.93\pm0.16$ for Pal 5
($r\approx19\kpc$). The GD--1 constraint is weaker because the stream
is located much closer to the disk; the Pal 5 constraint is especially
stringent because at its $(R,Z) \approx (8.4,16.8)\kpc$ location, the
potential is almost entirely dominated by the dark--matter halo and
the overall potential flattening that we measure ($q_\Phi =
0.94\pm0.05$) is almost directly that of the halo. While the estimates
are consistent, the radial trend is towards a more oblate halo further
out.

The combined constraint from Pal 5 and GD--1 is $c/a=
1.05\pm0.14$. This measurement of the halo axis ratio is obtained in
the context of a three-component potential model for the Milky Way,
but displays no correlations with the other parameters describing the
potential. This result is at odds with the misaligned, triaxial halo
at $20\kpc < r < 60\kpc$ found to describe the kinematics of the Sgr
stream well \citep{Law10a}. From numerical simulations including the
effect of growing a massive baryonic disk, we expect the Milky Way to
have $c/a\approx0.8$, which is in $\approx 2\sigma$ tension with our
measurement. While the tension is too mild currently to strongly
disfavor the standard collisionless cold dark matter model for the
formation of dark matter halos, our result hints at the importance of
processes that sphericalize the Milky Way halo.

We also measure the total mass of the halo within 20 kpc to be
$M_{\mathrm{halo}}(r<20\kpc) = 1.1\pm0.1\times10^{11}\msun$ with a
scale radius of $18.0\pm7.5\kpc$. Such a halo mass and scale radius
are fully consistent with the expected mass and concentration for the
Milky Way's halo.

The available data on tidal streams will soon increase in both
quantity and quality with the release of the \emph{Gaia}
data. Especially the precise, sub $\mas\yr\inv$ proper motion
measurements for the brightest stream members will significantly
improve the measurements of the local force from these streams (see
\figurename~\ref{fig:pal5samples} and
\figurename~\ref{fig:gd1samples}). It will also increase the amount of
data on other, more distant streams like the Orphan stream and
potentially wholly new streams. The methodology of the B14 stream
model as used in this paper will then allow fast, flexible fits to all
of the different streams that can be combined to further build up our
direct measurements of the force field as in
\figurename~\ref{fig:forcefield} and be interpreted in terms of disk
and halo models. Because steady--state modeling of the smooth halo
will also allow tight constraints on the gravitational forces in the
same volume, a comparison between the results from steady--state
modeling---sensitive to the current forces---and those from
streams---affected by the forces over the entire lifetime of the
stream---may also allow direct inferences about the growth of the
Milky Way's dark--matter halo (see, \eg, \citealt{Buist15a}).

All code used in this paper is made publicly available at
\\ \centerline{\url{http://github.com/jobovy/mwhalo-shape-2016}~.}

\acknowledgements We thank Naomi McClure-Griffiths for providing the
new HI terminal velocity measurements from \citet{McClure16a} in
electronic form and the anonymous referee for a constructive
report. JB and AB received support from the Natural Sciences and
Engineering Research Council of Canada. JB also received partial
support from an Alfred P. Sloan Fellowship. TF \& NK are supported by
the NSF CAREER award 1455260. The MCMC analyses in this work were run
using \emph{emcee} \citep{Foreman13a}.

\end{document}